\documentstyle[11pt,newpaspekg,twoside]{article}
\def\edcomment#1{\iffalse\marginpar{\raggedright\sl#1\/}\else\relax\fi}
\reversemarginpar

\begin{document}
\title{A Dwarf Galaxy Survey in the Local Volume}
 \author{Eva K.\ Grebel}
\affil{Dept.\ of Astronomy, U.\ of Washington, Seattle, WA 98195, USA\\
Hubble Fellow\\
MPI f\"ur Astronomie, K\"onigstuhl 71, D-69117 Heidelberg, Germany}
\author{Patrick Seitzer}
\affil{Dept.\ of Astronomy, U.\ of Michigan, Ann Arbor, MI 48109, USA}
\author{Andrew E. Dolphin}
\affil{Kitt Peak National Obs., NOAO, Tucson, AZ 85726, USA}
\author{Doug Geisler}
\affil{Dept.\ de Fisica, Grupo de Astronomia, U.\ de Concepci\'on, Chile}
\author{Puragra Guhathakurta}
\affil{UCO/Lick Observatory, UCSC, Santa Cruz, CA 95064, USA}
\author{Paul W. Hodge}
\affil{Dept.\ of Astronomy, U.\ of Washington, Seattle, WA 98195, USA}
\author{Igor D. Karachentsev}
\affil{Special Astrophysical Observatory, N.Arkhyz, KChR, 357147, Russia}
\author{Valentina E. Karachentseva}
\affil{Astron.\ Obs.\ of Kiev University, 04053, Observatorna 3, Kiev, Ukraine}
\author{Ata Sarajedini}
\affil{Astronomy Dept., Wesleyan University, Middletown, CT 06459, USA}

\begin{abstract}
We are carrying out a comprehensive multi-wavelength study of dwarf galaxies
in the Local Volume ($\le 5$ Mpc).  After our all-sky survey more than doubled 
the dwarf census we are now measuring structural parameters, 
integrated properties, and velocities.  Our 200--orbit HST 
snapshot program yields stellar content, star
formation history, and TRGB distances.  We can thus study the 
morphology--density relation in galaxy groups in three dimensions, 
the influence of environment on galaxy evolution, and the 
metallicity--surface brightness--luminosity relation.
\end{abstract}

\section{Introduction}

The ``Local Volume'' describes our extragalactic neighborhood out to
$\sim 5$ Mpc and includes galaxies with velocities $\le 500$ km s$^{-1}$.  
Apart from the Local Group, this volume contains several nearby galaxy 
groups as well as isolated field galaxies.  The galaxy census becomes
increasingly incomplete for smaller galaxies with low surface brightness,
particularly for dwarf galaxies.  Furthermore, while most dwarf galaxies 
in the Local Group have been studied in great detail, very little is known 
about the properties of dwarf galaxies beyond its boundaries.

For these reasons our group began a comprehensive study of dwarf galaxies 
in the Local Volume.  Karachentseva \& Karachentsev carried out careful
searches of photographic plates (POSS, ESO-SERC) of 97\% of the sky for new 
dwarf galaxy candidates and more than doubled the previously known census 
(e.g., Karachentseva \& Karachentsev 1998; 2000; Karachentseva et al.\ 1999;
2000).  

Ground-based verification and follow-up observations are currently being
carried out involving southern and northern hemisphere telescopes at 
APO, BTA, CTIO, Keck, and MDM. The imaging observations allow us to determine 
the degree of resolution of the dwarf galaxies as a first measure of distance, 
to estimate their morphological type, and to determine structural parameters
and integrated properties.  Radio observations (Huchtmeier et al.\ 2000)
tell us about their H\,{\sc i} content and radial velocity.  Ongoing optical
spectroscopy is used to measure velocities and abundances.

\section{Our HST snapshot survey}

Detailed optical follow-up of our dwarf candidates is being pursued with 
WFPC2 aboard HST.  We were awarded a total of 200 
orbits for a snapshot survey of nearby dwarf galaxy candidates; one galaxy
per orbit.  Our 
targets include known and possible dwarfs in nearby galaxy groups and in 
the field,  but also a few spirals and other massive 
galaxies to establish distances and relative positions within galaxy groups.  
Most of these galaxies are being studied for the first time. 

While we could in principle obtain data for 200 galaxies,
in practice the number will be lower due to the constraints in
scheduling snapshot observations.  At the time of
the present workshop in March 2000 we had already obtained data for 37 
galaxies, while 8 orbits were lost due to guide star acquisition failures
and due to the gyro failure in late 1999.  

We observe each target in two filters ($F606W$ and $F814W$) corresponding
to $V$ and $I$.  The non-standard $F606W$ filter was chosen over $F555W$
due to its greater sensitivity.  The exposure time in each filter is 600 s,
which allows us to reach $I \sim 25$ mag.  Our data are automatically 
processed and reduced with the photometry package {\em HSTphot} developed
by Dolphin (2000).     

\subsection{Resolved stellar populations}

Due to the short exposure times our data typically reach only the upper
portion of the red giant branch for galaxies beyond the Local Group.  Owing 
to the superior resolution of HST, however, these 10 min exposures with a
2.4-m space telescope are about 1 mag deeper than our 15 min exposures 
obtained with the 10-m Keck II telescope (Grebel \& Guhathakurta 1999).   

In dwarf galaxies with young populations our color-magnitude diagrams
(CMDs) show clearly the young main sequence and supergiants.  The locus of 
red supergiants and young asymptotic giant branch (AGB) stars can be strongly 
contaminated by Galactic foreground stars.  The blue main sequence and 
blue He-burning stars allow us to trace ages ranging from a few million 
years to several 100 Myr.  Where present, the horizontally tilted region of 
intermediate-age AGB stars just above the red giant branch (RGB) indicates 
populations older than $\sim 1$ Gyr.  The RGB itself may comprise both 
intermediate-age and old ($>10$ Gyr) stars.  The foreground reddening 
can be derived from the extinction maps of Schlegel et al.\ (1998).

While we obtain detailed age information for young populations, our age
resolution is poor for intermediate-age populations.
We cannot unambiguously detect old populations
due to the age degeneracy along the RGB and lack of deeper data that reach
at least the horizontal branch.  
Nonetheless the data on stellar content are invaluable for constraining the star
formation history and far supersede the information
previously available from ground-based imaging.

\subsection{Distances and metallicities}
 
The tip of the RGB (TRGB) allows us to derive distances for our dwarfs.  
For the dwarf spheroidal (dSph) galaxies in our sample the location of 
the TRGB can be determined to
$\sim \pm 0.1$ mag.  The main sources of uncertainty are 
photometric blends, crowding, or, in some cases, sparsely populated RGBs.
In dwarf irregulars (dIrrs) the determination of the TRGB magnitude is
complicated by the contamination of the RGB stars with younger populations
such as AGB stars.  We estimate the distance uncertainty to be of the 
order of $\pm 0.2$ mag. 
 
In heavily extincted galaxies of any type the apparent
magnitude of the TRGB may be too close to the detection limit to allow us
to derive anything else but a lower limit for the distance.  This is the
case for many of the members of the obscured IC342/Maffei group, while 
we are getting well-defined TRGBs for galaxies in groups such as Sculptor,
Centaurus A, and M81.
 
Despite their drawbacks TRGB distances are the best available distance
indicator in galaxies where Cepheids may be absent and RR Lyrae or the
horizontal branch are close to or beyond the detection limit of most 
present-day telescopes.  They provide an efficient means to derive distances
for a large sample of targets.  An interesting additional benefit is that
these distances are helpful in calibrating the density fluctuation method
recently developed for dSphs and dwarf ellipticals by Jerjen et al.\ (2000)
since we have several dwarfs in common. 

For those of our targets that suffer little foreground extinction and that
are closer than $\sim 3$ Mpc the slope and mean color of the upper RGB can 
be used to estimate mean metallicities and to constrain the metallicity 
spread.  

The mean metallicities derived this way, together with central surface
brightness and absolute magnitude, follow the same trend defined by
dwarf galaxies in the Local Group (Grebel \& Guhathakurta
1999, Karachentsev et al.\ 2000).

\subsection{Star clusters}

Our survey shows a few extended, spherical objects that resemble globular
clusters.  Applying color selections and comparison with the known colors
of Galactic globular clusters allows us to identify potential globular
cluster candidates and to estimate the specific globular cluster frequency.

Some of our targets appear to be nucleated.  In other cases compact objects 
are being identified as background galaxies based on their morphological
appearance and/or colors.
Concentrations of blue stars are seen in several of the dIrrs in our sample
and may represent young open clusters or OB associations.

\section{Summary}

Our Local Volume survey helps to improve the census of dwarf galaxies 
and to augment the faint end of the galaxy luminosity function.  Our
ground-based follow-up studies in optical and radio wavelengths yield
structural parameters, integrated properties, morphological types,
kinematic information, and an assessment of the gas content of the 
dwarf galaxy candidates.  Our HST snapshot survey yields, for the
first time, detailed information on the resolved stellar populations
and a crude assessment of their star formation histories.

The TRGB distances allow us to derive probable group membership and the 
location of galaxies within galaxy groups.  We thus obtain 
the three-dimensional structure of nearby groups.  This together
with our partial knowledge of the star formation episodes as well as 
the kinematic information helps us to understand the impact of environment
on dwarf galaxy evolution, allowing a three-dimensional study of the 
morphology-density relation.

\acknowledgments

Support for this work was provided by NASA through grant GO--08192.97A from
STScI,  which is operated by AURA, Inc., under NASA contract
NAS5--26555. EKG gratefully acknowledges support by NASA through grant
HF--01108.01--98A from the Space Telescope Science Institute. 
EKG, IDK, and VEK were also supported through an 
Henri Chr\'etien International Research Grant administered by the American
Astronomical Society.

\end{document}